\documentstyle[emulateapj,onecolfloat]{article}
\newcommand{\tskip}{}%\vspace{2pt}
\input epsf
\begin{document}
\def\etal{{\it et al.~}}

\twocolumn[
 %\submitted{Draft- \today}
\submitted{Accepted for Publication in the Astrophysical Journal}

\title{Measurement of Cosmic Shear with the Space Telescope Imaging
  Spectrograph}

\author{Jason Rhodes$^{1,2,3}$, Alexandre Refregier$^{4,5,3}$, Nicholas
R. Collins$^{1,6}$, Jonathan P. Gardner$^{1}$, Edward J.
Groth$^{7}$ \& Robert S. Hill$^{1,6}$} \affil{1 Laboratory for
Astronomy and Solar Physics, Code 681, Goddard Space
  Flight Center,  Greenbelt, MD 20771; rhodes@astro.caltech.edu}
\affil{2 NASA/National Research Council Research Associate}
\affil{3 California Institute of Technology, 1201 E California Blvd.,
  Pasadena, CA 91125}
\affil{4 Service d'Astrophysique, CEA Saclay, 91191
  Gif sur Yvette, France}
\affil{5 Institute of Astronomy, Madingley Road, Cambridge, CB3 OHA, U.K.}
\affil{6 Science Systems and Applications, Incorporated}
\affil{7 Department of Physics, Princeton University, Jadwin Hall,
  P.O. Box 708, Princeton, NJ 08544}

\begin{abstract}
Weak lensing by large-scale structure allows a direct measure of
the dark matter distribution.  We have used parallel images taken
with the Space Telescope Imaging Spectrograph (STIS) on the Hubble
Space Telescope to measure weak lensing, or cosmic shear. We
measure the shapes of 26036 galaxies in 1292 STIS fields and
measure the shear variance at a scale of 0.51 arcminutes. The
charge transfer efficiency (CTE) of STIS has degraded over time
and introduces a spurious ellipticity into galaxy shapes during
the readout process.  We correct for this effect as a function of
signal to noise and CCD position.  We further show that the
detected cosmic shear signal is nearly constant in time over the
approximately four years of observation. We detect cosmic shear at
the 5.1$\sigma$ level, and our measurement of the shear variance
is consistent with theoretical predictions in a $\Lambda$CDM
universe. This provides a measure of the normalization of the mass
power spectrum $\sigma_8=(1.02 \pm 0.16) (0.3/\Omega_m)^{0.46}
(0.21/\Gamma)^{0.18}$. The one-$\sigma$ error includes noise,
cosmic variance, systematics and the redshift uncertainty of the
source galaxies. This is consistent with previous cosmic shear
measurements, but tends to favor those with a high value of
$\sigma_8$. It is also consistent with the recent determination of
$\sigma_8$ from the Wilkinson Microwave Anisotropy Probe (WMAP)
experiment.

\end{abstract}
 ]

\section{Introduction}
\label{introduction} Weak gravitational lensing by large-scale
structure has become an important tool in understanding the amount
and distribution of dark matter (see Mellier \etal 2002 \&
Refregier 2003 for reviews). Most cosmic shear measurements have
been performed using wide-field ground-based telescopes. However,
the measurement of galaxy shapes from the ground is limited by
atmospheric seeing, while the \emph{Hubble Space Telescope (HST)}
has a much smaller Point Spread Function (PSF) ($<0.1''$ as
opposed to $\sim0.8''$ for ground-based telescopes). Several
groups have thus used \emph{HST} images to study cosmic shear
(Rhodes, Refregier \& Groth 2000 [RRGI] and 2001 [RRGII];
Refregier, Rhodes \& Groth 2002 [RRGIII]; H\"{a}mmerle \etal
2002). Recent measurements of cosmic shear from the ground (Bacon
et al., 2002, Hoekstra \etal 2002, \& van Waerbeke \etal 2002;
Jarvis \etal 2003; Brown \etal 2003) and from space (RRGIII) have
been used to derive an estimate of the normalization $\sigma_{8}$
of the matter power spectrum with an uncertainty comparable to
that of more traditional methods such as the measurement of
cluster abundances (see RRGIII for a discussion).

In this paper, we present the results of a weak lensing analysis
of images taken with the Space Telescope Imaging Spectrograph
(STIS) on \emph{HST}. The STIS data we use have been described by
Pirzkal et al. (2001). H\"{a}mmerle, et al. (2002) have presented
an analysis of a subset (121 of 1292 fields) of these data,
reporting a 4\% RMS shear at the $1.5\sigma$ level. Because there
are a number of different methods for measuring shear and
correcting for systematic effects, it is beneficial to compare the
results we present here to the results of other groups examining
data using the same instrument.

This paper is organized as follows.  In \S\ref{data} we describe the
STIS data used in this study. We detail the STIS parallel archive
maintained by the STIS Investigation Definition Team (IDT) at Goddard
Space Flight Center, including the methods used to reduce and catalog
STIS images. \S\ref{procedure} describes the procedure used to measure
the weak lensing shear in the STIS data, including the method for
removal of systematic effects. In \S\ref{results}, we present the
results, including constraints on cosmological parameters. We draw our
conclusions and briefly comment on the future of weak lensing
measurements in \S\ref{conclusions}.

\section{Data}
\label{data}
The STIS images used in this study were taken primarily in parallel
mode, meaning that another instrument on \emph{HST} was the prime
observing instrument. Thus, the images are essentially randomly
positioned on the sky, separated from the primary observation by 5 to
8 arcminutes depending on the primary instrument. This is ideal for a
study of cosmic shear because many random pointings on the sky
minimize the error due to cosmic variance.
%We exclude fields with galactic
%latitude $|b|<20^{\circ}$ because such fields suffer from stellar
%contamination.

There are four stages to the data reduction: (i) input image
co-alignment for a particular field, (ii) preliminary image
reduction that involves bias and dark subtraction, hot pixel
correction, flat-field division and correction for geometric
distortion, (iii) cosmic-ray (CR) removal and image combination,
and (iv) object detection.
%see /home/bhill/sci/stispar/aux/group_db_geom.cln
%    /home/bhill/sci/stispar/comb_onepass.pro
%    /home/bhill/sci/stispar/precr_par2.pro
%    /home/bhill/mypro/hotterp.pro
%    /home/bhill/sci/stispar/cr_reject.pro
%    /home/bhill/sci/stispar/aux/stisccd.sex

Step (i) involves determining the relative offsets using sources
identified in dithered images. Copies of the input images are
pre-cleaned to minimize the effect of CRs on the source
identification step. Small clusters of pixels with high data
values relative to surrounding pixels are first identified then
corrected using interpolation.  This pre-cleaning may modify or
remove real sources, but these image copies are used only for the
purpose of co-alignment. Source Extractor (\texttt{SExtractor})
(Bertin \& Arnouts 1996) is used to identify sources in the
pre-cleaned image copies. Those sources in common on all images
are used to find individual image offsets.  The average offset is
subtracted from the individual offsets so that all images will be
shifted to a common field center.

For step (ii), we used the Interactive Data Language (IDL) version
of CALSTIS (Lindler 2003), which was developed by the STIS
Instrument Definition Team (IDT).  A one-dimensional fit is made
to the CCD overscan region and subtracted from each column of the
image array to account for temporal variations between the
observation and the two-dimensional bias image.  A two-dimensional
bias frame is then subtracted from the science image.
%Bad pixels are flagged in a data-quality array.
Next, the dark image subtraction is performed and hot pixels are
corrected. The hot pixel tables from the Space Telescope Science
Institute archive list new hot-pixels created by the constant
on-orbit CR flux since the last dark image was obtained. For a
given observation, the hot pixels corrected are those in common
between the two lists made closest in time before and after the
observation. Hot pixels with a dark rate greater than 0.02
counts/second were corrected by linear interpolation among the
eight adjacent pixels. The resultant image is divided by two
flat-field images: one that accounts for pixel-to-pixel
variations, and another that corrects the low-frequency spatial
variations across the field of view due to vignetting.  A final
hot-pixel correction is performed to pre-clean the data before
cosmic-ray rejection.  Using a separate IDL program, small
clusters of pixels ($\leq$ 3-pixels wide) with high data values
relative to surrounding pixels are first identified then corrected
using interpolation.  This procedure is especially useful in
removing hot or bad pixels that would not otherwise be eliminated
in the CR-rejection step when there is no dither.  This is the
same algorithm described in step (i) for removing cosmic-rays
before running SExtractor, but in this case, the threshold for
correction is set to a higher value so that valid source pixels
are not modified. The correction for detector/optics induced shear
(or geometric distortion) was performed with bi-linear
interpolation using the values given by Malumuth and Bowers
(1997). The distortion coefficients were derived using the
astrometric shifts of stellar images.  All images are then shifted
to a common field center using bi-linear interpolation for the
sub-pixel offsets derived in step (i).

Once the individual images have been co-aligned they are combined
with the routine IDL \texttt{cr\_reject.pro}
%, available fromtheIDL Astronomy User's Library
%(http://idlastro.gsfc.nasa.gov/homepage.html)
(step iii). This program emulates the STSDAS
task\\ \texttt{stsdas.hst\_calib.wfpc.crrej} %\texttt{crrej}
and is equipped to handle input images with different exposure
times. The main difference between the two algorithms is in the
initial step of sky background determination. The procedure
\texttt{stsdas.hst\_calib.wfpc.crrej} uses the modal value of all
of the image pixels, while \texttt{cr\_reject.pro} uses an
estimate of modal value of a subset of image pixels.  Using a
subset of image pixels reduces the bias in the pixel distribution
caused by foreground source flux. The estimate uses an algorithm
from \texttt{DAOPHOT} I (Stetson, 1987). The average and standard
deviation of the pixel-value distribution are computed, and
outliers are removed using an iterative sigma-clipping method. If
the distribution is Gaussian (uncontaminated by foreground
sources) the mean, median and mode should be the same. If, after
20 iterations, the mean and median are the same, this value is
taken as the sky for the image. If the mean of the distribution is
larger than the median (a non-Gaussian distribution) the true sky
is estimated as 3$\times$median - 2$\times$mean.  The scalar
background so derived is subtracted from each input image. Pixels
are rejected as cosmic-rays if their value is greater than an
input number of $\sigma$ from a reference value. The constant read
noise, the statistical noise (square-root of the counts), and a
noise proportional to the counts comprise the $\sigma$.  This last
noise component accounts for differing point-spread-functions
(PSFs) or sub-pixel image shifts. It is expressed by the
``mult\_noise'' variable in \texttt{cr\_reject.pro} that we set to
0.03. If there are 7 or more input images, the first guess at a
CR-free or reference image is the pixel-by-pixel median of all the
input images, otherwise the minimum is used. The first pixel
rejection pass cuts all pixels greater than 8$\sigma$ from this
reference image. A second pass uses the mean of the cleaned images
as the reference image and a rejection criterion of 6$\sigma$. The
third and final pass uses a 4$\sigma$ rejection criterion on the
mean of the cleaned images from the second pass. The sky values
that were subtracted from each input image before CR-rejection are
added back into the final set of cleaned images. The final image
combination is the weighted average of the cleaned images. The
weight parameters are the sky noise and the read noise. The final
image pixels are in units of counts per second.

For step (iv) we use \texttt{SExtractor} (Bertin \& Arnouts 1996)
for object detection, photometry, star/galaxy separation and to
compute positions for each detected object. \texttt{SExtractor}
subtracts a smoothed two-dimensional background with a mesh size
of 1$\arcsec$ (\texttt{BACK\_SIZE}$=20$) and filter size setting
of \texttt{BACK\_FILTERSIZE}$=3$. The image is then convolved with
a 1$\farcs$25 x 1$\farcs$25 0$\farcs$5-FWHM Gaussian filter.  The
detection and analysis thresholds are both set to 0.7$\sigma$
above the measured sky noise.  We require that 5 contiguous pixels
meet this threshold criterion to be considered a valid source.  In
order to separate blended sources, we set the number of deblending
sub-thresholds to 32, and the minimum contrast parameter for
deblending to 0.005. The input stellar seeing disk FWHM, used for
star-galaxy separation, is 0$\farcs$07.
%The gain is set to account for the gain of the detector (GAIN=1),
%and for the exposure time of the observation, so that
%the image statistics
The photometry reported is taken from the MAG\_AUTO column of the
output catalog. MAG\_AUTO is the source magnitude measured inside
a unique elliptical aperture for every object (Kron 1980). The
two-element input parameter \texttt{PHOT\_AUTOPARAMS} controls the
elliptical apertures.  The first element is the k-factor described
in Bertin \& Arnouts (1996), and the second is the minimum
possible radius (in pixels) for an elliptical, or Kron, aperture.
We use the default values of 2.5 and 3.5, respectively, for the
two elements. A 12-pixel annulus outside the elliptical aperture
is used to determine the local sky background around each object.

We examined 2335 fields imaged in 21 \emph{HST} programs (7781,
7782, 7783, 7908,7910, 7911, 8062, 8064, 8084, 8091, 8393, 8406,
8470, 8545, 8549, 8562, 8796, 8808, 8870, 8884, \& 9248). Several
of these programs are described elsewhere: the STIS Parallel
Survey (SPS; Gardner et al 1998; Teplitz \etal 2003a \& 2003b)
obtained filterless imaging and slitless spectra of random fields
on the sky and program 8562 (PI Schneider; continued as 9248) was
conducted to gather data specifically to search for cosmic shear
(Pirzkal et al. 2001). Of these fields, 1494 pass a visual
inspection that indicates they are not contaminated by stars or
large objects. In our measurement of cosmic shear, we use 1292 of
these fields that pass several other cuts described below.

\begin{figure}[h]
\plotone{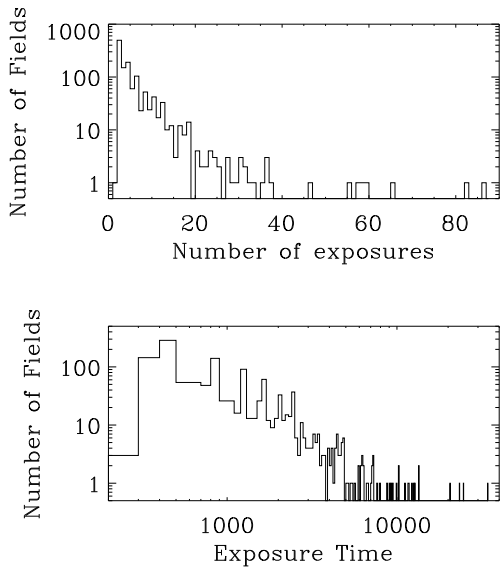} \caption{Top: Histogram of the number of
individual exposures going into each image.  Bottom: Histogram of
the exposure time for each individual image.}
\label{fig:exposures}
\end{figure}

The selected fields were imaged between August 9, 1997 and May 15,
2001.  We do not use more recent data because of possible changes
in systematic effects that may have occurred when STIS was forced
to go to backup electronics as a result of a malfunction in July
2001 (Brown 2001). Each field has between 2 and 86 individual
exposures with total exposure times ranging from 265 to 34,200
seconds. The distribution of exposure times and number of
exposures are shown in Figure~\ref{fig:exposures}.

Imaging is done using the 50CCD clear mode, a very wide bandpass
sensitive to light in the wavelength range $220<\lambda<1100$nm
and limited by the sensitivity of the CCD (Baum et al. 1996). We
co-add exposures that are within 5 arcseconds of each other. Only
exposures taken within 6 months of each other are co-added to
produce the final image. This limit is chosen to avoid co-adding
exposures with large differences in their calibration files.

%Each exposure is bias and dark subtracted and flat fielded.  We
%use SExtractor (Bertin \& Arnouts 1996) for object detection,
%photometry, star/galaxy separation and to compute positions for
%each detected object. A more detailed description of the image
%reduction and object catalog generation, as well as the
%photometric zero point $ZP_{50CCD}=26.52$ is given in Gardner
%\etal (1998) .

\begin{figure}[h]
\plotone{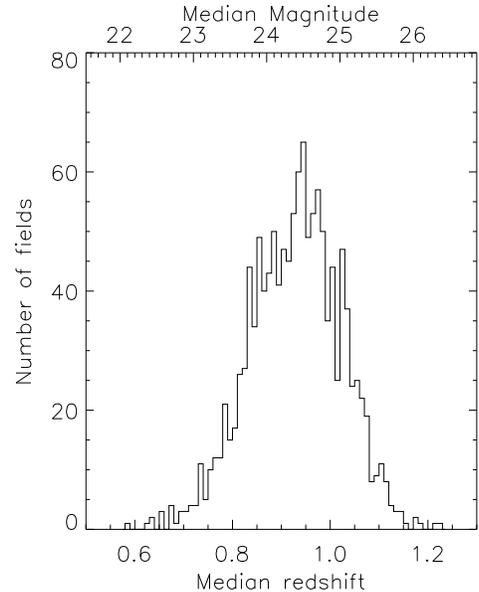}
 \caption{Distribution of the median redshifts of
the 1292 fields we use for cosmic shear measurements.  The
redshift $z_m$ is estimated from the median magnitude of the
selected galaxies in the field via Equation~(\ref{eq:z}).}
\label{fig:zhist}
\end{figure}

Using the results of RRG III and the fact that $\langle M \rangle
\approx \langle I \rangle +1$  for typical galaxy colors where $M$
and $I$ are the STIS and WFPC2 magnitude respectively, we find
that the median magnitude $M_{m}$ of the galaxies in each STIS
field is related to their median redshift $z_{m}$ by
\begin{equation}
\label{eq:z}
%z_{m} \approx 0.722 + 0.149 \times( M_{m} - 23.0 )
z_{m} \approx 0.72 + 0.15 \times( M_{m} - 23.0 )
\end{equation}
We show a histogram of the median redshifts of the selected fields
in Figure~\ref{fig:zhist}.

We visually examined 2335 fields in the STIS archive. Our selection
criteria were: two or more exposures, unbinned pixels (to maintain the
full STIS resolution), and imaging performed with the clear 50CCD
filter. We discarded fields that have a very bright saturated star,
fields that are crowded by stars (common for fields with galactic
latitude $|b|<20^{\circ}$), and fields with a large object that
covered a substantial portion of the field. This left us with 1494
fields.  Because of repeat visits by the \emph{HST} separated by more
than 6 months and visits to fields that are separated by only of order
tens of arcseconds, some of these fields overlap.  So as not to double
count fields, we discarded one of any pair of fields that lie closer
than $1'$ to each other (the STIS Field of view is
$50''\times50''$). When considering which field to discard, we kept
the field with the greater total exposure time.

Due to several reasons including thermal effects, the \emph{HST}
undergoes jitter, or shaking (Nelan \& Makidon 2002). This jitter
tends to blur images and introduces changes into measured object
shapes. The median root mean square (RMS) jitter in one direction
is 3-4 mas, but the RMS jitter value can be greater than 100 mas.
We thus discard fields in which any of the input exposures had an
RMS jitter greater than 10 mas (0.2 STIS pixels) in either of two
orthogonal directions, or for which no jitter information is
available.  After these cuts, we were left with 1292 fields.

\section{Procedure}
\label{procedure}
\subsection{Shape Measurement}
We follow the method described in RRGI to measure the object
shapes and sizes and to correct for systematic effects. We measure
the second $I_{kl}$ and fourth $I_{klmn}$ order moments for each
object. These moments are defined as a sum over pixels $p$ given
by, for example,
\begin{equation}
\label{eq:moment}
I_{kl}=\frac{\sum_{p} x_k x_l I({\bf x})
w({\bf x})}{\sum_{p}I({\bf x})w({\bf x})}
\end{equation}
where $I({\bf x})$ is the intensity in a pixel with position
${\bf x}=(x_1,x_2)=(x,y)$ with respect to the
centroid of the object and $w({\bf x})$ is a Gaussian weighting
function. Similar equations hold for the remaining moments. The
standard deviation of the weight function is chosen to be (in pixels)
\begin{equation}\label{eq:gauss}
\sigma_{w}=\textrm{max}\left[3,\sqrt{\frac{A}{\pi}}\right],
\end{equation}
where the area $A$ is determined from the SExtractor measured
semi-major($a$) and semi-minor($b$) axes as
$A=\pi(\frac{a+b}{2})^{2}$. The minimum weight function size of 3
pixels (0.15 arcseconds) is the optimal weight function we found for
point sources (stellar images).  This is similar to the minimum weight
function size of 2 pixels (0.2 arcseconds) found in RRGI for
WFPC2.

These measured moments are used to compute the two component
ellipticity of each galaxy given by
\begin{equation}
\label{eq:e}
e_1=\frac{I_{11}-I_{22}}{I_{11}+I_{22}},~~
e_2=\frac{2I_{12}}{I_{11}+I_{22}}.
\end{equation}
The first component of this
ellipticity ($e_1$) indicates elongation along the x (positive $e_1$)
and the y (negative $e_1$) axes.  $e_2$ indicates elongation along
axes $45^{\circ}$ and $-45^{\circ}$ from the x axis. We also define the
RMS size $d$ of the object as
\begin{equation}
\label{eq:d}
d^2=\frac{1}{2}(I_{11}+I_{22}).
\end{equation}

There are several systematic effects which affect the measured
shapes of galaxies. The PSF of the telescope introduces smearing
that can be deconvolved into an isotropic and an anisotropic
component. The optics introduce an instrumental shear into the
galaxies. Our method to correct for these effects is presented in
RRGI. Here, we describe the particular steps taken for the STIS
data set.

%\subsection{Correction for Geometric Distortion}

\subsection{PSF Model}
\label{psfmodel} 
\begin{figure}[h]
\plotone{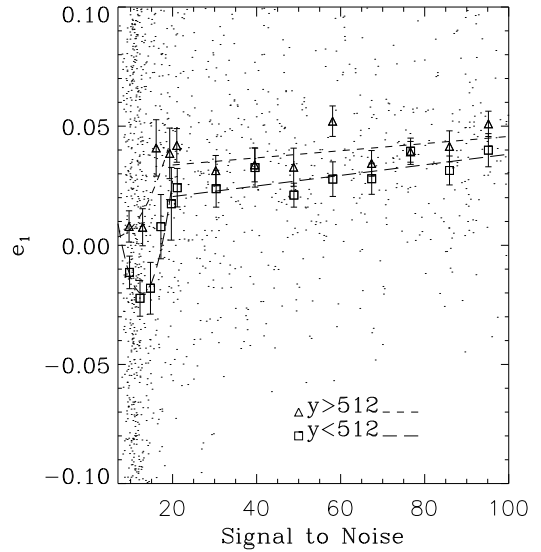}

\caption{Stellar ellipticity $e_1$ as a function of signal to noise
(S/N) for stars in galaxy-dominated fields. The triangles are binned
averages in the top half of the chip, squares are binned averages in
the bottom half of the chip. The error bars represent 1 $\sigma$
errors in the mean. The dashed lines represent linear fits for
$20<\textrm{S/N}<100$ and quadratic fits for $\textrm{S/N}<20$.  These
fits are used to derive corrections for the second order moments
$I_{kl}$ based on S/N and $y$ position.}
\label{fig:sn_e1}
\end{figure}

Ground-based surveys typically use stars in the
survey fields to measure and correct for the PSF. Due to the small
field of view of \emph{HST}, space-based surveys do not have
enough stars per field to do such a correction and have relied on
separate observations of high signal-to-noise (S/N) stellar fields
to measure the PSF (see e.g. RRGI; RRGIII; H\"{a}mmerle et al
2002). However, in examining the STIS PSF, we found that the PSF
shape depends on the S/N.  As shown in Figure~\ref{fig:sn_e1}, the
average ellipticity ($e_1$) of stars in the survey fields changes
from negative (elongation along the y axis) to positive
(elongation along the x axis) as S/N increases. This result is
consistent with a previously known charge transfer efficiency
(CTE) effect in STIS (Kimble, Goudfrooij, \& Gilliland 2000).
Objects with few counts (due to short exposure times or low S/N)
bleed in the $y$ direction causing an elongation of the object in
that direction.  This CTE effect has been shown to be worse in the
($y<512$)half (hereafter the ``bottom half'') of the STIS CCD
(Woodgate 2002).  Figure~\ref{fig:sn_e1} confirms that the
elongation in the $y$ direction is indeed more pronounced in the
bottom half of the STIS CCD, providing further evidence that the
CTE is indeed the source of the S/N dependence of the PSF.

The CTE effect described above complicates the modelling of the PSF,
and thus the correction of galaxy shapes for PSF smearing, for two
reasons.  The first is that we cannot simply measure high S/N stellar
PSFs and use these to correct low S/N galaxies.  The second is that we
cannot use separate stellar fields to measure the PSF.  Stars in
stellar dominated fields (such as images of globular clusters) do not
suffer the same CTE effects as stars in sparsely populated fields,
even at low S/N.  This is because the higher background levels in
stellar fields fill in the charge traps causing the CTE effect.  Thus,
low S/N stars in high-background fields have a PSF similar to the PSF
of high S/N stars regardless of background. Since the bulk of our
galaxies are low S/N galaxies in low-background fields, that is the
regime in which we must model the PSF.

For S/N$>100$ we see no evidence for an $e_1$ dependence on either S/N
or $y$ position.  Therefore, we consider the high S/N PSF to be the
``true'' STIS PSF and model the S/N and $y$ dependence of the PSF as
perturbations about this true PSF.  We select 534 high ($>100$) S/N
stars from our survey and fit the 8 stellar moments from these stars
to 5th order polynomials in $x$ and $y$ position as described in
RRGI. Since we are attempting only a first order correction for the
effects of S/N and $y$ position, we do not correct the fourth order
moments for these effects. The low S/N moment at any position $(x,y)$
is given by the high $S/N$ moment at that position, as calculated from
the 5th order polynomial fits, minus some correction factor.  The
correction factor is the difference between the average value of the
moment at high S/N and the average value of the moment at low S/N at
the $y$ position of the star on the chip.  Thus,
\begin{equation}\label{eq:corr}
I_{kk}^{low}(x,y)=I_{kk}^{high}(x,y)-
(\overline{I_{kk}^{high}}-\overline{I_{kk}^{low}}(y,\textrm{S/N})),
\end{equation}
for each diagonal component $k=1,2$.

We then write the moments as
\begin{equation}
\label{eq:ixx}
I_{11}=(1+e_1)d^2,~~
I_{22}=(1-e_1)d^2
\end{equation}
where the size $d$ is defined as in equation~(\ref{eq:d}). This
size is slightly dependent on S/N, but has no dependence on $y$
position. The size $d^2$ as a function of S/N is well fit by a
linear equation given in Table~\ref{tab:fits}. We fit $e_1$ to the
S/N in two different bins of $y$ position, $y<512$ and $y>512$. We
use linear fits for the range $20<\textrm{S/N}<100$ and quadratic
fits for the range $8<\textrm{S/N}<20$. Table~\ref{tab:fits} gives
the best fit functions, which are plotted on
Figure~\ref{fig:sn_e1}.  To calculate the average value of $e_1$
at a particular S/N and $y$, we perform a linear interpolation or
extrapolation between the two $y$ fits at that S/N value assuming
that the $y<512$ fit is valid for $y=256$ and the $y>512$ fit is
valid for $y=768$. We find that there is no S/N or $y$ dependence
for $e_2$, which is consistent with zero. Thus, no $I_{12}$
correction is necessary. These fits allow us to calculate the
corrections to the high S/N stellar moments using
Equation~(\ref{eq:corr}).

The average stellar ellipticity we measure at high S/N (S/N$>100$),
3-5\%, is slightly higher than the 1-2\% measured by H\"{a}mmerle et
al (2002).  We compare our measurement with theirs by obtaining one of
the high S/N stellar fields (o48b41010\_3\_ass) in their publicly
available catalog and performing our analysis on that image. This
image is drizzled to increase resolution and thus has a linear pixel
size of 0.025'' (half of the intrinsic STIS pixel size). If we analyze
this image using the method outlined above with a Gaussian weight
function width $\sigma=6$ pixels (twice what we use in our
non-drizzled images), we find an average stellar ellipticity
$e_1=4-5$\%.  This is stable for larger weight function widths, but if
we use a smaller weight function width ($\sigma=3$ drizzled pixels) we
measure an average $e_1=1.5$\%. Thus, we are able to recover the value
measured by H\"{a}mmerle et al (2002) by using a sufficiently compact
weight function.  The weight function we use to measure stars
represents a trade-off between the reduction of noise in moment
measurements and the measurement of the full PSF anisotropy.  As in
RRGI, stellar moments are relatively stable for small perturbations in
this weight function width.

\subsection{Correction for the PSF}
Using the stellar moments--corrected for weighting and detector shear
as described in RRGI and fit to S/N and $y$ position as described
above--we can correct the galaxy moments for the effects of PSF.  The
first step is a correction for the anisotropic smearing of the PSF and
the second step is a correction for the isotropic portion of the PSF
which increases the galaxy size slightly. This correction method was
specifically tuned to \emph{HST} images with small PSF and has
produced excellent results in the correction of WFPC2 images for PSF
effects (RRGII; RRGIII).  The corrected galaxy moments are used to
calculate the ellipticity of each galaxy, which allows the calculation
of the overall shear in each field as described below.

\subsection{Shear Measurement}
 We detected 63,895 objects in the 1292
fields we used. We discard stars and objects for which no jitter
information is available or for which the jitter is very high. We
further discard objects which have spurious ellipticities
(non-physical post-correction ellipticities usually caused by poor
background subtraction), leaving us with 49692 objects. We then
discard objects that are too small to allow a shape measurement,
i.e. with $d<1.7$ pixels (after corrections), leaving 35033
objects.  In RRGII and RRGIII, a magnitude cut was made to discard
galaxies too faint to allow for shape measurement. Here, we opt to
use a \emph{S/N} cut based on SExtractor values of flux and flux
error. A cutoff value of $S/N=10$ left us with 26036 galaxies.

We thus obtain the corrected ellipticity $e_{i}$ of each selected
galaxy that passed all the cuts mentioned above.  We then
calculate the shear estimator $\gamma_i$ for each galaxy using the
equation
\begin{equation}
\gamma_i=\frac{e_i}{\langle G \rangle}
\end{equation}
where the shear susceptibility factor $\langle G \rangle \simeq 1.4$
is calculated according to Equation~28 of RRGI and averaged over all
galaxies in our final sample.

\section{Results}
\subsection{Shear Measurement and Cosmological Parameter
Estimation} \label{results} To quantify the cosmic shear
statistics, we follow the procedure of RRG III. The cosmic shear
variance $\sigma_{\rm lens}^{2}$ is obtained from a weighted sum
of the estimators $\sigma_{{\rm lens},f}^{2}$ for each field $f$
as
\begin{equation}
\sigma_{\rm lens}^{2} = \sum_{f} w_{f} \sigma_{{\rm lens},f}^{2},
\end{equation}
where the weights $w_{f} \propto \sigma_{{\rm noise},f}^{-4}$ are
normalized so that $\sum_{f} w_{f} \equiv 1$.
%$\sigma_{{\rm
%noise},f}^2$ is the shear variance arising from noise for field
%$f$ and can be measured.
 We define the noise variance $\sigma_{{\rm
noise},f}^2\equiv\langle|\gamma_{f}^{noise}|^2\rangle$, which is
measured from the data by computing the error in the mean
$\gamma_{if}$ from the distribution of galaxy shears in each field
(RRGIII).  This provides a nearly optimal weighting scheme which
allows us to use all fields that are not contaminated by stars or
bright objects. Fields with  few objects that pass our selection
criteria are appropriately down-weighted.

With our STIS data set, we find
\begin{equation}
\sigma_{\rm lens}^{2} = (5.43 \pm 1.06 \pm 1.74) \times 10^{-4},
\end{equation}
where the first error corresponds to noise only, while the second
includes noise, cosmic variance and systematics. Our detection of
cosmic shear (first error) is thus significant at the $5.1\sigma$
level. The systematics error is dominated by the uncertainty in
our PSF correction (see \S\ref{procedure}). To estimate it, we
first noted that the uncertainty in our measurement of the PSF
ellipticity $e_{*}$ in figure~\ref{fig:sn_e1} is $\Delta e_{*}
\simeq 0.01$. The resulting uncertainty in the galaxy shears is
$\Delta \gamma \simeq G^{-1} (\langle d_{*} \rangle / \langle d
\rangle)^2 \Delta e_{*} \simeq 1.0\times10^{-3}$ (see RRG I,
equation~58), where the average rms size (see Eq.~[\ref{eq:d}]) of
the PSF (with S/N=10, see table~\ref{tab:fits}) and of galaxies
(after our cuts) is $\langle d_{*} \rangle \simeq 1.2$ pixels and
$\langle d \rangle \simeq 3.2$ pixels, respectively.

\begin{figure}[h]
\plotone{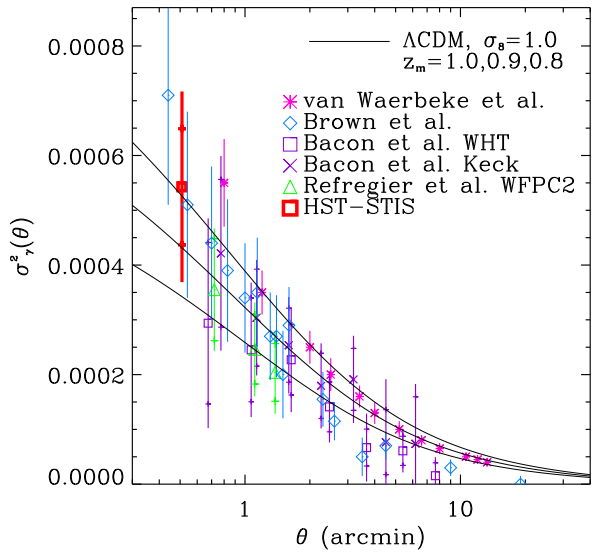} \caption{Cosmic shear variance $\sigma_{\rm
lens}^{2}$ as a function of the radius $\theta$ of a circular
aperture. Our observed value (HST-STIS) is shown as well as that
from other recent measurements: van Waerbeke et al. (2001), Bacon
et al.  (2002, WHT and Keck), Brown et al. (2003), Refregier et
al.  (2002, RRGIII, HST-WFPC2). Also displayed are the predictions
for a $\Lambda$CDM model with $\Omega_{m}=0.3$, $\Gamma=0.21$,
$\sigma_{8}=0.7$ and $1.0$, and a range of galaxy median redshift
$z_{m}=0.8,0.9$ and $1.0$, using the Smith et al. (2003) nonlinear
fitting function. When relevant, the inner error bars include
statistical errors only, while the outer error bars include
statistical errors and cosmic variance.} \label{fig:var}
\end{figure}

Our measurement of the shear variance is shown in
figure~\ref{fig:var}.  Recent measurements from other surveys with
similar galaxy redshifts ($z_{m} \simeq 0.8$--$1$) are also shown for
comparison (RRGIII; van Waerbeke \etal 2002; Brown \etal 2003; Bacon
\etal 2002) are also shown for comparison. The solid lines show the
predictions for a $\Lambda$CDM model with $\Omega_{m}=0.3$,
$\sigma_{8}=1$, and $\Gamma=0.21$ and with the above range of galaxy
median redshift $z_{m}$. Our variance measurement is consistent with
these other measurements and with the $\Lambda$CDM prediction. It also
agrees with the measurement of H\"{a}mmerle et al. (2002) who found
$\sigma_{\rm lens}^2 \simeq (15\pm12)\times 10^{-4}$ (from their
figure 15). Note that their error bar is considerably larger than
ours, because their measurement is based on about $1/10$ of the
fields we have used.

Our measurement allows us to set constraints on $\sigma_{8}$, the
amplitude of matter fluctuations on $8 h^{-1}$ Mpc scales. In a
$\Lambda$CDM model with $\Omega_{m}=0.3$ and $\Gamma=0.21$ the
shear variance is given by
\begin{equation}
\sigma_{\rm lens}^2 \simeq 5.17 \times10^{-4}
  \left( \frac{\sigma_{8}}{1.0} \right)^{2.42}
  \left( \frac{\Omega_m}{0.3} \right)^{1.10}
  \left( \frac{\Gamma}{0.21} \right)^{0.44}
  \left( \frac{z_m}{1.0} \right)^{1.85},
\end{equation}
where $z_{m}$ is the median redshift of the galaxies, whose
redshift distribution was assumed to be $p(z) \propto z^{2}
e^{-(z/z_0)^2}$ with $z_0=z_m/1.09$. The scale $0.51'$ is the
effective radius of a circular cell corresponding to the chip size
($50''$ on a side). The details of this calculation can be found
in Bacon \etal (2000). Note however, that unlike these authors, we
have used the more accurate fitting function of Smith \etal (2003)
to compute the non-linear corrections to the power spectrum.

For our sample and weighting scheme, the effective median magnitude of
the galaxies is $M_{m} \simeq 24.8 $. According to
Equation~\ref{eq:z}, this corresponds to a median redshift of
$z_{m}=1.0\pm0.1$, where the $1\sigma$ error arises from the
uncertainty in this equation. Our measurement of $\sigma_{\rm lens}^2$
thus yields
\begin{equation}
\sigma_{8}=(1.02 \pm 0.16) \left( \frac{0.3}{\Omega_m} \right)^{0.46}
  \left( \frac{0.21}{\Gamma} \right)^{0.18}
\end{equation}
where the error includes noise, cosmic variance, systematics, and the
redshift uncertainty. This result is consistent at the $1\sigma$ level
with some earlier cosmic shear surveys (Bacon et al. 2002; RRGIII; van
Waerbeke et al. 2002) which derived values of $\sigma_{8}$ in the
range 0.9--1.0 for $\Omega_{m}=0.3$ (see figure~9 and table~1 in
Refregier 2003 for a summary of the latest cosmic shear results). It
is also consistent, but only at the $<2\sigma$ level, with other
cosmic shear surveys which found $\sigma_{8}$ in the range 0.7--0.8
for the same value of $\Omega_{m}$ (Brown et al. 2003; Hamana et
al. 2003; Jarvis et al. 2002; Hoekstra et al. 2002). The recent WMAP
results (Spergel et al. 2003) yield $\sigma_{8}=0.91\pm0.21
(0.3/\Omega_m)^{0.6}$ when taken alone (for their single power law
model), and $\sigma_{8}=0.78^{+0.08}_{-0.10}(0.3/\Omega_m)^{0.6}$ when
combined with other CMB and large-scale structure data (for their
running power law index model). Our results are thus also consistent
with these WMAP values at the $<1.5\sigma$ level.

\subsection{Temporal Stability of the Shear Measurement}
The CTE effect described  in section \S\ref{psfmodel} is growing
worse as the STIS CCD ages (Goodfrooij \&  Kimble 2002).  In order
to test whether this degradation affects our results, we perform
two tests.  The first test involves stars and the second test
involves examining our final result for temporal stability.

There are sufficient stars in the survey to perform the analysis
described in \S\ref{psfmodel} on two subsections of the data.  We
divide the data roughly in half by time, classifying data taken
before mid 1999 as ``early'' and data taken after that time as
``late.''  We find that for S/N$>20$ there is little change to the
fit to $e_{1}$ shown in Figure~\ref{fig:sn_e1}. However, for
S/N$<20$ we find that the fits shown in Figure~\ref{fig:sn_e1} do
change slightly. The early data are fit by curves with $e_1$
slightly higher than those shown in Figure~\ref{fig:sn_e1} and the
late data show a slightly lower $e_1$.  This is consistent with a
CTE that is growing worse over time.  In both cases, the departure
from the curve shown in Figure~\ref{fig:sn_e1} is about 1\%. This
average  1\% systematic error in the correction of stellar
ellipticities at low S/N is included in the error estimates for
our values of $\sigma^{2}_{\gamma}$ and $\sigma_{8}$ (see
\S\ref{results}).

\begin{figure}[h]
\plotone{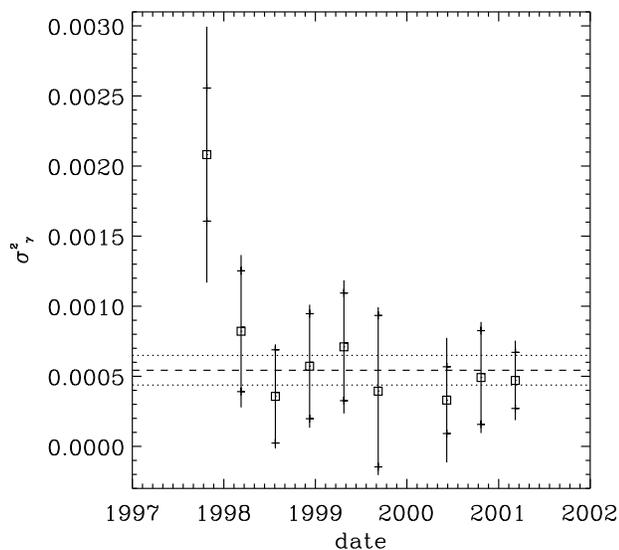} \caption{Cosmic shear variance $\sigma_{\rm
lens}^{2}$ as a function of the date. The inner error bars include
statistical errors only, while the outer error bars include
statistical errors and cosmic variance. The dotted horizontal
lines represent the value of $\sigma_{\rm lens}^{2}$ and the
associated one sigma statistical error for the entire survey. The
left-most point is consistent with what H\"{a}mmerle \etal (2002)
found using roughly the same data.} \label{fig:dates_shear}
\end{figure}

To perform the other test, we divide our galaxy data into nine
equally spaced date bins. We then measure the RMS shear
$\sigma^{2}_{\gamma}$ in each bin. We show these measurements in
Figure~\ref{fig:dates_shear}. This plot shows no trend over time
as might be expected if the CTE effect was not being sufficiently
corrected for. The shear signal is consistent throughout the time
in which the data we use in this paper was taken.  The only
marginal outlier is the very first bin, which is $1.7\sigma$  from
the value for the entire survey when statistical errors and cosmic
variance are included. We have run several tests to determine why
data taken during this period gives a higher shear signal than
data taken during the rest of the survey but find nothing special
about the data in this bin. Our results in this date range are
consistent with those of H\"{a}mmerle \etal (2002) who found a
high shear signal using STIS data from roughly the same period.
%Furthermore, a single bin out of nine $1.7\sigma$ away from the
%mean is expected for gaussian statistics.
Excluding the first bin (by only analyzing data taken after
December 1997) only changes the measured $\sigma^{2}_{\gamma}$
from $5.43\pm1.74\times10^{-4}$ to $4.76\pm1.68 \times10^{-4}$.
The effect on $\sigma_8$ is less pronounced; the removal of the
first bin decreases $\sigma_8$ from $1.02\pm0.16$ to
$0.97\pm0.16$.

\section{Conclusions}
\label{conclusions}

We have measured the shapes of 26036 galaxies in 1292 STIS fields,
corresponding to about 0.25 square degrees.  We corrected for
systematic effects as outlined in RRGI with the added step of
correcting galaxy moments for a S/N dependent charge transfer
efficiency. We have detected a cosmic shear signal using the STIS
parallel archive at the $5.1\sigma$ level.  After correcting the
galaxy shapes for PSF distortions, detector shear and CTE effects,
we measure a shear variance of $\sigma_{\rm
lens}=(5.43\pm1.74)\times10^{-4}$ on the STIS chip scale ($50''$
on a side), where the $1\sigma$ error includes noise, cosmic
variance and systematics. This is consistent with the earlier and
much noisier measurement of H\"{a}mmerle et al. (2002). For a
$\Lambda$CDM model with $\Omega_m=0.3$ and $\Gamma=0.21$, this
sets a constraint on the amplitude of the matter power spectrum
given by  $\sigma_{8}=1.02 \pm 0.16$, where the $1\sigma$ error
includes noise, cosmic variance, systematics and redshift error.
This is consistent with earlier measurements of $\sigma_{8}$ from
cosmic shear, but tends to favor those with higher values. It is
also consistent with the recent determination of $\sigma_8$
(Spergel et al. 2003) from CMB anisotropies with the WMAP mission.

The results presented here represent a contribution to the first
generation of space-based weak lensing results presented in RRGII,
RRGIII, \& H\"{a}mmerle \etal (2002). The next generation of
space-based parallel observations optimized for weak lensing will
be made with the Advanced Camera for Surveys (ACS) on the
\emph{HST}.  A targeted ACS project, the COSMOS 2 square degree
field (GO-9822) will provide strong constraints on cosmological
parameters and allow for high resolution dark matter map. ACS
surveys will benefit from the high resolution of ACS (0.05
arcsecond pixels like STIS) as well as its enlarged area
(approximately 10 square arcminutes) and improved sensitivity
(relative to WFPC2).  New methods for the measurement of object
shapes and the correction of systematic effects are being
developed that will capitalize on the excellent resolution of this
survey and future space-based surveys (Refregier \& Bacon 2003;
Bernstein \& Jarvis 2002). Future generations of weak lensing
surveys both from the ground (e.g., the Canada-France-Hawaii
Legacy survey, Mellier \etal 2000; the Large-aperture Synoptic
Survey Telescope; Tyson \etal 2003) and from space (e.g. the
Supernova Acceleration Probe; Alcock \etal 2003; Rhodes \etal
2003, Massey \etal 2003; Refregier \etal 2003) will continue to
utilize this unique method to measure cosmological parameters to
unprecedented accuracy. These projects will survey large areas
while obtaining images in multiple filters allowing for accurate
photometric redshifts of the survey galaxies. SNAP, with its
near-infrared capability, high resolution, and wide field of view,
will be able to probe mass concentrations beyond $z=1$, thus
allowing for redshift tomography and the study of the growth of
structure.  This will enable weak lensing to make a contribution
to the study of dark matter and dark energy, and to set tight
constraints on cosmological parameters.

\acknowledgments We thank Richard Massey and Richard Ellis for
useful discussions. We thank the referee Peter Schneider for
useful comments and suggestions.  AR was supported in Cambridge by
an EEC fellowship from the TMR network on Gravitational Lensing
and by a Wolfson College Research Fellowship.  EJG was supported
by NASA Grant NAG5-6279.  JR was supported by an National Research
Council-GSFC Research Associateship. We thank Bruce Woodgate,
Elliot Malumuth, Randy Kimble, Ted Gull, Terry Beck, Keith Feggans
and the rest of the STIS team at GSFC for their support in
cataloging and understanding the STIS parallel images.

\newpage
\begin{table*}[t]
 \begin{center}
\caption{PSF Model Parameters for Equations 4 and 5}
\label{tab:fits}
 \nopagebreak
 %\tableline
\begin{tabular}{|p{2cm}|p{2cm}|p{2cm}|p{7cm}|}

 \tableline

Parameter & S/N range & $y$ range & Fit\\
\tskip\tableline\tableline\tskip%\footnotesize
 $d^2$ & 10-100 & all &$1.47+0.00072\times \textrm{S/N}$\\
$e_1$ & 8-20 & $y<512$
&$0.124-0.023\times\textrm{S/N}+0.00089\times(\textrm{S/N})^2$\\
$e_1$&8-20 & $y>512$
&$0.010-0.0025\times\textrm{S/N}+0.00021\times(\textrm{S/N})^2$ \\
$e_1$& 20-100&$y<512$ &$0.016+0.0022\times\textrm{S/N}$\\
$e_1$ & 20-100 & $y>512$&$0.031+0.00015\times\textrm{S/N}$ \\
$e_2$ & all & all & 0\\
  \tableline
\end{tabular}
\end{center}

\end{table*}
\normalsize

\end{document}